\newcommand{\diracslash}[1]{#1\llap{/\kern2pt}}

\def\bearr{\begin{eqnarray}}
\def\eearr{\end{eqnarray}}

\newcommand{\be}{\begin{equation}}
\newcommand{\ee}{\end{equation}}
\newcommand{\bea}{\begin{eqnarray}}
\newcommand{\eea}{\end{eqnarray}}
\newcommand{\ba}[1]{\begin{array}{#1}}
\newcommand{\ea}{\end{array}}

\documentclass[prd,
twocolumn,
aps,floats,nofootinbib,tightenlines,showpacs]{revtex4}
\usepackage{graphicx}
\usepackage{epsfig}
\usepackage{amsmath}
\begin{document}
\title{Natural inflation at the GUT scale}
\author{
Subhendra Mohanty and  Akhilesh Nautiyal }
\affiliation{Theory Division, Physical Research Laboratory,
Navrangpura, Ahmedabad 380 009, India}
\date{\today}

\begin{abstract}
Natural inflation driven by pseudo-Nambu-Goldstone bosons have a problem that the nearly scale invariant spectrum of density perturbations is attained   only
when the symmetry breaking scale is of the order of Planck scale. 
We show here
that if one couples the PNGB to a thermal bath as in warm inflation models,
the amplitude and spectral index which agrees with the Wilkinson 
Microwave Anisotropy Probe (WMAP) data is obtained
with the symmetry breaking in the GUT scale. We give a GUT model of PNGB arising
out of spontaneously broken lepton number at the GUT scale which gives rise to
heavy Majorana masses for the right handed neutrinos which is needed in see-saw
models. This model also generates a lepton asymmetry because of the derivative
coupling of the PNGB to the lepton current. A characteristic feature of this
model is the prediction of large non-gaussianity which may be observed in the
forthcoming PLANCK experiment.
\end{abstract}

\maketitle

\section{introduction}
Inflation \cite{Guth} was introduced to solve the horizon and curvature problems of cosmology and in addition it predicted a scale invariant spectrum of density perturbations which was verified by the Cosmic
Background Explorer (COBE), WMAP and other CMBR anistropy experiments.The successful model of inflation requires a flat potential and a natural candidate for such a potential is the Pseudo-Nambu-Goldstone potential as first pointed out in \cite{natural1,natural2,natural3}.

One limitation of natural inflation models is that the symmetry breaking scale $f$ is related to the spectral index $n_s =1-M_p^2/(8 \pi f^2)$ and observations of microwave anisotropy constrain the symmetry breaking scale to be close to the Planck scale. As discussed in Banks et al \cite{Banks:2003sx} a symmetry breaking scale larger than $M_P$ makes the theory susceptible to large quantum corrections which can destabilize the flat PNGB potential. There have been several
attempts at solving this large $f$ problem in natural inflation. Arkadi-Hamed
et al \cite{ArkaniHamed:2003wu} invoke extra dimensions with the Wilson loop of a gauge field in the extra dimension to explain why $f \sim M_P$. Similar arguments are also given by Kaplan and Weiner
\cite{Kaplan:2003aj}. Kim et al \cite{Kim:2004rp} invoke two field natural 
inflation to bring down the symmetry breaking scale below Planck scale. Kinney 
and Mahanthappa \cite{Kinney:1995cc} show that in some special symmetry 
breaking schemes  the quadratic term in the PNGB field is subdominant compared 
to the higher order terms  and in these models the symmetry breaking scales 
can be lower than the Planck 
scale.

In this paper we show that if the PNGB inflaton of the natural inflation model is coupled to a radiation bath (with a sub-dominant energy density) as in warm
inflation models \cite{warm} the symmetry breaking scale $f$ can be in
the GUT scale and be consistent with the observations of the temperature
anisotropy spectrum observed by WMAP \cite{Komatsu:2008hk}. In this model the
dissipative coupling of the PNGB inflaton makes it roll slowly even in a steep
potential which results when $f$ is lowered from $M_P$ to $M_{GUT} \sim 10^{16} GeV$.

As a specific model let us consider the SU(5) model where the right handed neutrino $N$ is a singlet. In the see-saw mechanism \cite{see-saw} one generates a heavy Majorana mass by coupling this right handed neutrino to a SU(5) singlet Higgs,
\be
-\mathcal L_{\nu}=g H N N^C.
\label{yukawa}
\ee
In order to break lepton number spontaneously we have a
potential for the Higgs
\be
-\mathcal L_H=\frac{\lambda}{8}(H^{\dagger}H-\frac{f^2}{2})^2.
\label{vhiggs}
\ee
Here $f$ is the spontanous symmetry breaking scale.

At the minima of the potential the Higgs is given by $H=\frac{1}{\sqrt 2}\,
f\, e^{i\frac{\phi}{f}}$. Here angular variable $\phi$ is the Goldstone boson
of the spontaneously broken lepton number symmetry.
Quantum gravity effects are expected to break global symmetries at the Planck scale.
If there is an explicit symmetry breaking due to gravity the Goldstone boson
acquires mass.
The explicit symmetry breaking term can be of the form
\be
-\mathcal L=\frac{M^2}{M_P}N N^C+O(\frac{1}{M_P^2}).
\label{esb}
\ee
Because of this explicit symmetry breaking the potential of PNGB is given by \cite{Frieman}
\be
V(\phi)=\Lambda^4\left(1+\cos\left(\frac{\phi}{f}\right)\right).
\label{vpngb}
\ee
$\Lambda$ is related to explicit symmetry breaking scale  $\mu=\frac{M^2}{M_P}$. The mass of the PNGB is given by $m_\phi=\frac{\mu^2}{f}=\frac{\Lambda^2}{f}$. This implies that $\Lambda=\mu=\frac{M^2}{M_P}$. Now if we take
$M\sim\, M_{GUT}\sim\, 10^{16}-10^{17}$GeV then we have $\Lambda\sim\,
10^{13}-10^{14}$GeV which is the allowed range by WMAP data.

\section{Warm natural inflation}

In warm inflation the  equation of motion of inflaton field is given by
\be
\ddot{\phi}+(3H+\Gamma)\dot{\phi}+V^{\prime}(\phi,T)=0.
\label{inflaton}
\ee
  Here ${V^\prime}$ denotes differentiation of $V$ with respect to $\phi$, $\Gamma$ is
the damping term and $V(\phi,T)$ is thermodynamic potential.
 In slow roll approximation we neglect $\ddot{\phi}$ in the Eq.~(\ref{inflaton}). During inflation
the potential energy of the inflaton field dominates over radiation density. So
the dynamics of $\phi$ field is governed by
\bea
\dot{\phi}&=&-\frac{V^{\prime}}{3H+\Gamma}
\label{phidot},\\
H^2&=&\frac{8 \pi}{3 M_p^2}V.
\label{friedmann1}
\eea

The slow role parameters are
defined as
\bea
\epsilon&=&\frac{M_p^2}{16 \pi}{\left(\frac{V^{\prime}}{V}\right)}^2, \, \,
\eta=\frac{M_p^2}{8 \pi}\frac{V^{\prime\prime}}{V} \,\,, \nonumber\\
\beta&=&\frac{M_p^2}{8 \pi}\frac{\Gamma^{\prime}V^{\prime}}{\Gamma V},\, \,
\delta=\frac{M_p^2}{8 \pi}\frac{T V^{\prime}_{, T}}{V^{\prime}}.
\eea
   Here two extra slow roll parameters appear because of $\phi$ dependence of damping
term and temperature dependence of the potential.\\
The density perturbations during warm inflation are generated by thermal
fluctuations. The power spectrum for the density perturbations given in
\cite{Hall} is
\be
P_{\mathcal R}=\left(\frac{\pi}{4}\right)^{\frac12}\frac{H^{\frac52}
\Gamma^{\frac12}T}{\dot{\phi}^2}
\label{power}
\ee
which can be written in terms of potential and its derivative
using  Eq. (\ref{friedmann1}) and  Eq.
(\ref{phidot}) as
\be
P_{\mathcal R}=\left(\frac{\pi}{4}\right)^{1/2}\left(\frac{8\pi}{3 M_p^2}\right)^{5/4}
\frac{V^{5/4}\Gamma^{5/2}T}{V^{\prime 2}}.
\ee
 Using the natural inflation potential (\ref{vpngb}) we get for the power
spectrum,
\be
P_{\mathcal R}=\left(\frac{\pi}{4}\right)^{1/2}\left(\frac{8\pi}{3 M_p^2}\right)^{5/4}
\frac{\Gamma^{5/2}\, T\, f^2}{\Lambda^3}\frac{{\left(1+\cos\frac{\phi}{f}\right)}^{(5/4)}
}{\sin^2\frac{\phi}{f}}.
\ee

The spectral index can be defined as
\be
n_s-1=\frac{\partial \ln P_{\mathcal R}}{\partial \ln k}.
\label{ns1}
\ee
In terms of the slow roll parameters this can be written as
\be
n_s-1=\frac{3\, H}{\Gamma}\left(-\frac{9}{4}\epsilon+\frac{3}{2}\eta-\frac{9}{4}\beta \right).
\ee
For the given potential (\ref{vpngb}) the spectral index will be
\be
n_s-1=-\frac{3\, H}{\Gamma}\frac{3 M_p^2}{64 \pi f^2}
\frac{\left(3+\cos\frac{\phi}{f}\right)}{\left(1+\cos\frac{\phi}{f}\right)}.
\ee

\begin{figure}
\begin{center}
\epsfig{file=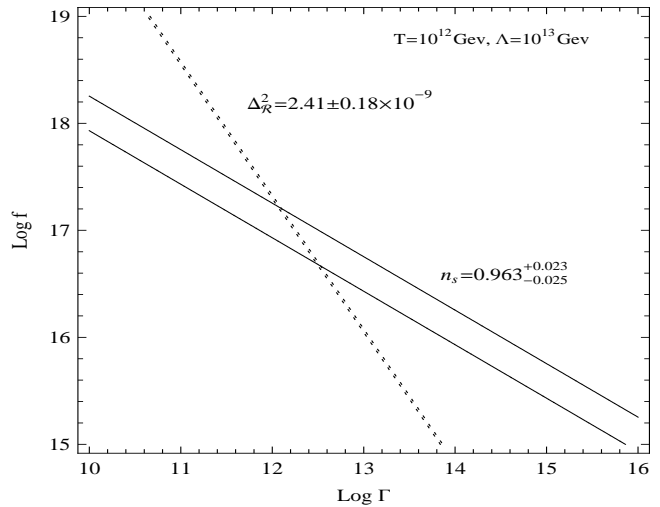,angle=0,width=9cm,height=7cm}
\end{center}
\caption []{The allowed range of $f$(GeV) and $\Gamma$(GeV) from the range of
 spectral index $n_s$ and the amlitude of curvature perturbations
$\Delta_{\mathcal R}^2$, from WMAP.}
\label{fig}
\end{figure}

  The observational constraint on $n_s$ from WMAP 5-year data \cite{Komatsu:2008hk}  is $0.948\, <\, n_s\, <\, 0.977$.
So it is obvious from above  Eq. that if we take warm inflation in strong
dissipative regime i.e $\Gamma$ is very large compared to $H$, we 
can have small value of $f$ (fig.~\ref{fig}).
In the cold natural inflation models on the other hand the spectral index
$n_s= 1-M_p^2/(8 \pi f^2)$. This implies that in the cold natural inflation
models WMAP data gives a strong constrain $f>0.7M_P$ \cite{Savage:2006tr}.

   The slow roll parameter $\epsilon$ for this model is
\be
\epsilon=\frac{M_p^2}{16 \pi f^2}\frac{\sin^2\frac{\phi}{f}}
{{\left(1+\cos\frac{\phi}{f}\right)}^2}.
\ee
  At the end of inflation $\epsilon=1+r$,  where $r=\frac{\Gamma}{3 H}$.
 This will give $\phi_f$ as
\be
\cos\frac{\phi_f}{f}=\left(\frac{1-\left(1+r\right)\frac{16 \pi f^2}{M_p^2}}{1+\left(1+r\right)\frac{16 \pi f^2}{M_p^2}}\right).
\ee
 Putting $r=3.9\times 10^4$ and $f=8\times10^{16}$GeV we get $\phi_f=2.9\, f$.
The e-foldings may be calculated as
\bea
  N &=&\int_{\phi_i}^{\phi_f} \frac{H}{\dot{\phi}}d\phi \,\,\,\,
    =\frac{8\, \pi\, \Gamma}{3H\, M_p^2}\int_{\phi_f}^{\phi_i}\frac{V}{V^{\prime}}d\phi \nonumber\\
    &=&\frac{16\, \pi\,  \Gamma\, f^2}{3H\, M_p^2}\left(\log\frac{\sin\left(\frac{\phi_f}{2f}\right)}{\sin\left(\frac{\phi_i}{2f}\right)}\right).
\eea
 The scalar field lies between $\pi f$ and $0$. For $N=60$ we get
$\phi_i=1.02f$. The value of the scalar field remains in the GUT regime and
still gives adequate e-foldings to solve the horizon and curvature problems.
\section{Microphysical model for large dissipation}
To get large dissipation the inflaton can be coupled  to another
scalar field $\chi$  by another explicit symmetry breaking term 
\be
\mathcal L_\chi=2 g^2 \phi^2\chi^2
\label{phichi}
\ee 
which in turn is
coupled to the radiation field $\sigma$ as 
\be 
\mathcal L_{\chi\sigma}= \frac{1}{\sqrt 2} h f\left(\sigma^2\chi^\star +
\chi^2\sigma^\star\right). 
\label{chisigma}
\ee 
This two step coupling
is necessary in order to generate a large dissipation without
destabilizing the inflaton potential by loop corrections
\cite{Berera:2008ar}.
 The dissipation
coefficient $\Gamma$ for this model has been calculated by Berera et
al \cite{Berera:2008ar}, 
\be
\Gamma=\frac{16}{\pi}\frac{g^2}{h^2}T\ln\frac{T}{m_{\chi}}
\label{gamma1}\ee 
The interaction terms in the Lagrangian
(\ref{phichi}) can generate one loop corrections to the
inflaton mass that can destabilize the flatness of the potential
(\ref{vpngb}). For the potential to remain flat the mass correction
$g^2\, f^2$ should be smaller than $\frac{\Lambda^4}{f^2}$. If we
take $\Lambda\sim 10^{13}$GeV and $f\sim 10^{16}$GeV then $g \le
10^{-6}$. For the validity of above expression (\ref{gamma1}), the
mass of $\chi$ field should be smaller than $T$. So if we take one
loop correction to the mass of $\chi$ field ($T\sim 10^{12}$GeV)
because of $\sigma$ field $h$ should be smaller than $10^{-4}$. If
we take $g$ and $h$ of the same order we can have $\Gamma \sim
10^{12}$Gev.
\section{Predictions for non-gaussianity}
Non-gaussianity is a very important characteristic of the model of
inflation. Its magnitude is conventionally defined by the parameter
called $f_{NL}$, which is the ratio of the three point correlation
to the two point correlation. In standard inflation non-gausianity
parameter $f_{NL}$ is proportional to the slow roll parameter and
are therefore small \cite{Gangui,Bartolo:2004if}. In warm inflation
models non-gaussianity arises because of non-linear coupling between
inflaton and radiation. The $f_{NL}$ for warm inflation models has
been calculated in \cite{Moss:2007cv}. It is given by \be
f_{NL}=-15\ln\left(1+\frac{\Gamma}{42\, H}\right)-\frac{5}{2}.
\label{fnl} \ee Taking the allowed range of $\Gamma$ from the fig
(\ref{fig}) i.e $1 \times 10^{12}\, <\, \Gamma\, <\, 3 \times
10^{12}$ we get $-122.6 \, <\, f_{NL}\, <\, -106.2$ which is allowed
by WMAP-5 data \cite{Komatsu:2008hk} ($-151\,<\, f_{Nl}\, <\, 253$).

\section{Leptogenesis}
 This model automatically generates lepton asymmetry at the end of inflation.
The PNGB coupling to lepton current is obtained from (\ref{yukawa}) as
\be
\mathcal L_{int}=\frac{1}{f}\partial_\mu\phi\, j^\mu_L
\ee
For the homogenous inflaton this will be
\be
\mathcal L_{int}=\frac{\dot\phi}{f} n_L
\ee

\begin{figure}
\begin{center}
\epsfig{file=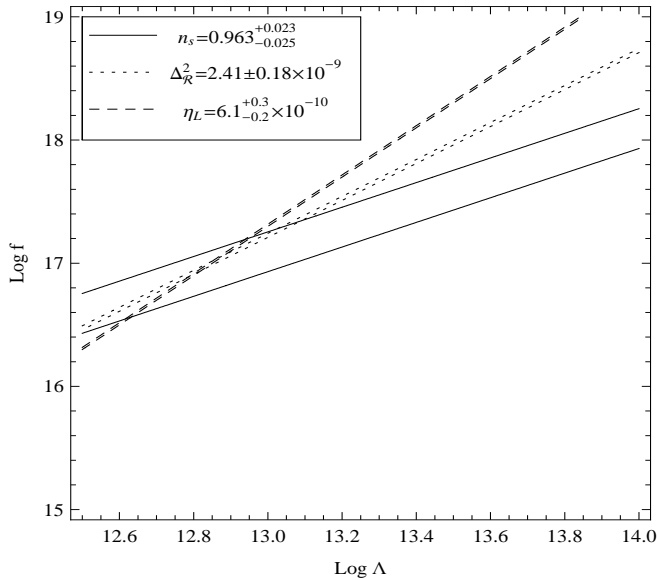,angle=0,width=9cm,height=8cm}
\end{center}
\caption []{The allowed range of $f$(GeV) and $\Lambda$(GeV) using spectral index $n_s$
curvature perturbations $\Delta_{\mathcal R}^2$ and lepton to entropy ratio
$\eta_L$for $T=10^{12}$GeV and $\Gamma=10^{12}$GeV}.
\label{fig2}
\end{figure}
here $n_L$ is lepton number. Therefore $\frac{\dot\phi}{f}$ is like a chemical
potential for the lepton number, $\mu_L=\frac{\dot\phi}{f}$. At equilibrium
the lepton number is given by
\bea
n_L&=&g_\nu\, \frac{T^3}{6}\left(\frac{\mu_L}{T}\right)\nonumber\\
&=&g_\nu\, \frac{\dot\phi\, T^2}{6f}.
\eea
So the lepton to entropy ratio will be
\be
\eta_L=\frac{n_L}{s}=\frac{15}{4\, \pi^2}\frac{g_\nu\, \dot\phi}{g_{\star}\, f\, T}.
\ee
Using slow roll approximation $\dot \phi=-\frac{V^{\prime}}{\Gamma}$.
For this model we get
\be
\eta_l=\frac{15}{4\, \pi^2}\frac{g_\nu\, \Lambda^4}{g_{\star}\, f^2\, \Gamma\, T}.
\label{nbys}
\ee
If we take $\Lambda\sim 10^{13}$GeV, $f\sim 10^{17}$GeV, $\Gamma\sim 10^{12}$
GeV and $T\sim 10^{12}$GeV,
 we get from (\ref{nbys}) $\eta_l\sim 10^{-10}$ (fig.~\ref{fig2}).

 If the lepton number is violated spontaneously at scale $f$ then there is an
effective lepton number violating dimension five operator \cite{Weinberg}
 \be
 \mathcal L_{\not L}= \frac{2}{f} h h l l + hc
 \label{barL}
 \ee
 where $l$ is the lepton doublet and $h$ is the Higgs doublet of the standard model. When the electroweak symmetry is broken by the Higgs acquiring a vev $ v$ then it generates a light neutrino mass
$ m_\nu= 4 \frac{ v^2}{f}$.
 The operator (\ref{barL}) can wipe out any generated lepton number at high temperature by the lepton number violating interactions $l +h \rightarrow l^c + h^\dagger$. The interaction rate of this lepton number violating reaction is \cite{Sarkar}
 \be
 \Gamma_{\not L} =0.04 \frac{ T^3}{f^2}.
 \ee
 These lepton number violating interactions will decouple at a temperature
 \be
 T_d=4.16\, \left(\frac{f^4\, \Lambda^4}{M_P^2} \right)^{\frac{1}{6}}.
\ee
For $\Lambda\sim 10^{13}$GeV and $f\sim 10^{17}$GeV this temperature is
$T\sim 10^{14}$GeV. Since the temperature of the radiation bath is
$T < 10^{13}$GeV the lepton asymmetry generated by the rolling PNGB field will not be
washed out by lepton number violating interactions with the light Higgs.

 The fact that PNGB's coupling to the lepton/baryon  current is of the derivative coupling form which gives rise to spontaneous leptogenesis of Cohen and
Kaplan \cite{cohen} was first recognized by Dolgov et al
\cite{Dolgov}. In \cite{Dolgov} a natural inflation without damping
was examined for generation of baryon/lepton number. It was found
that oscillations of the inflaton at the end of inflation wipes out
the baryon/lepton asymmetry so the PNGB model of creating B/L
asymmetry during natural inflation was considered unfeasible
\cite{Dolgov}. In \cite{Brandenberger} it was shown that if one
assumes the chaotic inflation potential $m^2 \phi^2$ and couples the
inflaton to radiation as in warm inflation and in addition assumes a
$\partial_\mu \phi j^{\mu}_{B,L}$ coupling of the inflaton then one
can get the required baryon asymmetry  with a suitable choice of
parameters.

\section{Conclusions}

There has been a long standing problem with utilizing the flat
potential of PNGB's for inflation as the nearly scale invariant power spectrum
which is consistent with observations generated only when the symmetry
breaking scale $f \sim M_P$ \cite{natural1,natural2,natural3,Banks:2003sx,ArkaniHamed:2003wu,Kaplan:2003aj,Kim:2004rp}. In this paper we show that by coupling
the inflaton to a radiation bath (as in warm inflation models \cite{warm}) can
reduce  $f$ to the GUT scale. The value of
the inflaton field $\phi \sim f\sim M_{GUT}$ which makes the inflaton potential
stable against Planck scale radiative corrections. We give a model of inflation
where the inflaton is the PNGB arising from spontaneous breaking of
lepton number which also gives a large Majorana mass for the right handed
neutrinos as required in see-saw models \cite{see-saw}. Since the 
PNGB's have a
derivative coupling to the
lepton current this model also generates a lepton asymmetry spontaneously
\cite{cohen}
during inflation. We show that with the parameters of the inflation model
which give the correct amplitude and spectral index of CMBR also give the
required lepton asymmetry of $\eta_L \sim 10^{-10}$ which can be converted to
a baryon asymmetry of the same
order by sphaleron processes in the electro-weak era \cite{Fukugita:1986hr}.


\begin{thebibliography}{M}
\bibitem{Guth}
  A.~H.~Guth,
  Phys.\ Rev.\  D {\bf 23}, 347 (1981).

  \bibitem{natural1}
  K.~Freese and W.~H.~Kinney,
  Phys.\ Rev.\ D {\bf 70}, 083512 (2004)
  [arXiv:hep-ph/0404012].
\bibitem{natural2}
  K.~Freese, J.~A.~Frieman and A.~V.~Olinto,
  %
  Phys.\ Rev.\ Lett.\  {\bf 65}, 3233 (1990).

\bibitem{natural3}
  F.~C.~Adams, J.~R.~Bond, K.~Freese, J.~A.~Frieman and A.~V.~Olinto,
  %
  Phys.\ Rev.\ D {\bf 47}, 426 (1993)
  [arXiv:hep-ph/9207245].
\bibitem{Banks:2003sx}
  T.~Banks, M.~Dine, P.~J.~Fox and E.~Gorbatov,
  JCAP {\bf 0306}, 001 (2003)
  [arXiv:hep-th/0303252].
\bibitem{ArkaniHamed:2003wu}
  N.~Arkani-Hamed, H.~C.~Cheng, P.~Creminelli and L.~Randall,
  Phys.\ Rev.\ Lett.\  {\bf 90}, 221302 (2003)
  [arXiv:hep-th/0301218].
\bibitem{Kaplan:2003aj}
  D.~E.~Kaplan and N.~J.~Weiner,
  JCAP {\bf 0402}, 005 (2004)
  [arXiv:hep-ph/0302014].
\bibitem{Kim:2004rp}
  J.~E.~Kim, H.~P.~Nilles and M.~Peloso,
  JCAP {\bf 0501}, 005 (2005)
  [arXiv:hep-ph/0409138].
\bibitem{Kinney:1995cc}
  W.~H.~Kinney and K.~T.~Mahanthappa,
  Phys.\ Rev.\  D {\bf 53}, 5455 (1996)
  [arXiv:hep-ph/9512241].

\bibitem{warm}
  A.~Berera,
  Phys.\ Rev.\ Lett.\  {\bf 75}, 3218 (1995)
  [arXiv:astro-ph/9509049].
\bibitem{Komatsu:2008hk}
  E.~Komatsu {\it et al.}  [WMAP Collaboration],
  arXiv:0803.0547 [astro-ph].
\bibitem{see-saw}
  R.~N.~Mohapatra and G.~Senjanovic,
  Phys.\ Rev.\ Lett.\  {\bf 44}, 912 (1980).

\bibitem{Frieman}
  J.~A.~Frieman, C.~T.~Hill, A.~Stebbins and I.~Waga,
  Phys.\ Rev.\ Lett.\  {\bf 75}, 2077 (1995)
  [arXiv:astro-ph/9505060].
\bibitem{Hall}
  L.~M.~H.~Hall, I.~G.~Moss and A.~Berera,
  Phys.\ Rev.\  D {\bf 69} (2004) 083525
  [arXiv:astro-ph/0305015].
\bibitem{Savage:2006tr}
  C.~Savage, K.~Freese and W.~H.~Kinney,
  Phys.\ Rev.\  D {\bf 74}, 123511 (2006)
  [arXiv:hep-ph/0609144].
\bibitem{Berera:2008ar}
  A.~Berera, I.~G.~Moss and R.~O.~Ramos,
  arXiv:0808.1855 [hep-ph].
\bibitem{Gangui}
  A.~Gangui, F.~Lucchin, S.~Matarrese and S.~Mollerach,
  Astrophys.\ J.\  {\bf 430}, 447 (1994)
  [arXiv:astro-ph/9312033].
\bibitem{Bartolo:2004if}
  N.~Bartolo, E.~Komatsu, S.~Matarrese and A.~Riotto,
  Phys.\ Rept.\  {\bf 402}, 103 (2004)
  [arXiv:astro-ph/0406398].
\bibitem{Moss:2007cv}
  I.~G.~Moss and C.~Xiong,
  JCAP {\bf 0704}, 007 (2007)
  [arXiv:astro-ph/0701302].

\bibitem{Weinberg}
  S.~Weinberg,
  Phys.\ Rev.\ Lett.\  {\bf 43}, 1566 (1979).

\bibitem{Sarkar}
  U.~Sarkar,
  arXiv:hep-ph/9809209.


\bibitem{cohen}
  A.~G.~Cohen and D.~B.~Kaplan,
  Phys.\ Lett.\  B {\bf 199} (1987) 251.
\bibitem{Dolgov}
  A.~Dolgov and K.~Freese,
  Phys.\ Rev.\  D {\bf 51}, 2693 (1995)
  [arXiv:hep-ph/9410346];\\
  A.~Dolgov, K.~Freese, R.~Rangarajan and M.~Srednicki,
  Phys.\ Rev.\  D {\bf 56}, 6155 (1997)
  [arXiv:hep-ph/9610405].

  \bibitem{Brandenberger}
  R.~H.~Brandenberger and M.~Yamaguchi,
  Phys.\ Rev.\  D {\bf 68}, 023505 (2003)
  [arXiv:hep-ph/0301270].
\bibitem{Fukugita:1986hr}
  M.~Fukugita and T.~Yanagida,
  Phys.\ Lett.\  B {\bf 174}, 45 (1986).


\end{thebibliography}
\end{document}